\documentclass[aps,prc,twocolumn,showpacs,nofootinbib,floatfix,groupedaddress]{revtex4-1}
\usepackage{graphicx}
\usepackage{dcolumn}
\usepackage{bm}
\usepackage{color}
\usepackage{amsmath}

\begin{document}

\title{Isospin Symmetry violation in mirror E1 transitions:\\
Coherent contributions 
from the Giant Isovector Monopole Resonance in $^{67}$As$-^{67}$Se}

\author{P.G. Bizzeti}
\affiliation{Dipartimento di Fisica e Astronomia, Universit\`a di Firenze
and INFN, Firenze, Italy}
\email{bizzeti@fi.infn.it}

\author{G. de Angelis}
\affiliation{Laboratori Nazionali di Legnaro, INFN, Legnaro (PD), Italy}

\author{S.M. Lenzi}
\affiliation{Dipartimento di Fisica e Astronomia, Universit\` a di Padova
and INFN, Padova, Italy}

\author{R. Orlandi}
\affiliation{IEM - CSIC, Madrid, Spain}

\date{\today}

\begin{abstract}
The assumption of an exact isospin symmetry would imply equal strengths
for mirror E1 transitions (at least, in the long-wavelength limit).
Actually, large violations of this symmetry rule { have been indicated} by a 
number of experimental results, the last of which is the 
$^{67}${As} -- $^{67}${Se} doublet investigated at GAMMASPHERE.
Here, we examine in detail various possible origins of the observed
asymmetry. The coherent effect of Coulomb-induced mixing with the
high-lying Giant Isovector Monopole Resonance
is proposed as the most probable process to produce  a large asymmetry in the 
E1 transitions,
with comparatively small effect on the other properties of the parent and 
daughter levels.
\end{abstract}

\keywords{21.10.Hw, 21.90.+f, 23.20.Js, 23.20.Lv}

\maketitle 

\section{\label{S:0}Introduction}
The presence of symmetries in physical laws in most cases greatly
enhances our understanding of their nature and consequences. 
Symmetries, either exact
or only approximate, have a particular importance in the fields of elementary
particle and nuclear physics. The approximate charge independence of nuclear 
forces, ultimately related to the near degeneracy of up and down quarks
\cite{Machleidt2001}, permits to treat protons and neutrons as different
states of the same particle (the nucleon) and to classify nuclear states
according to the different representation of a symmetry group, the Isospin
SU(2). In this scheme, protons and neutrons are characterised by the isospin
quantum number
$T=1/2$, with third component $T_3=+1/2$ and $-1/2$, respectively.
States of nuclei with the same mass number $A$ can be grouped,
according to the value of the isospin $T$,  in
isospin multiplets of $2T+1$ states belonging to the different nuclei,
distinguished by the value of $T_3=(Z-N)/2.$
{ Isospin}  symmetry is violated by the electromagnetic interaction
(mostly due to Coulomb forces among protons) and, to a lesser extent, also
by nuclear forces. However, the most important part of the
Coulomb interactions is diagonal with respect to $T_3$ and mainly 
contributes to the mass difference between various members of the
isospin multiplet.
Finer effects of the symmetry-breaking forces can be investigated by
measuring the so-called Mirror Energy Differences \cite{bentley2007}
or, more generally, differences in excitation energies
among members of a multiplet. { In recent years, this} field has become the object
of a considerable number of experimental and theoretical studies, as the level 
schemes of nuclei with $T_3=+1/2$ (i.e. Z=N+1)
could be measured for increasingly larger values of A. Furthermore, when
transition probabilities could be determined, their comparison
between mirror nuclei opened an important window to
investigate the amount and the origin of isospin violation.

\begin{figure}[t]
\includegraphics[width=\columnwidth,bb= 118 178 371 271]
{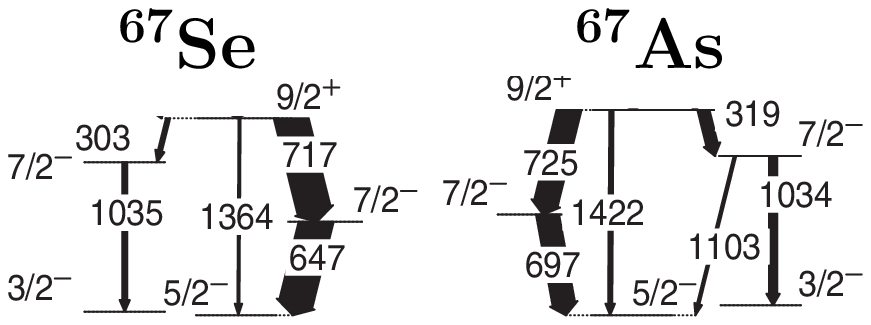}
\caption{\label{F:1}
(adapted from \cite{letter}) Partial level scheme of 
$^{67}$Se and $^{67}$As, showing the decay of the lowest $9/2^+$
state. Energy labels are in keV.}
\end{figure}

Here, we limit our discussion to the relatively simple case of E1 
transitions \cite{warburton69}.
The E1 transition operator is expected to be pure isovector, at least 
in the limit
of long wavelengths, where Siegert's theorem~\cite{siegert} 
holds. This fact implies that
(1) E1 transitions with $\Delta T=0$ in nuclei with $Z=N$ are forbidden, and 
that (2) corresponding E1 transitions in mirror nuclei have equal reduced strength.
{
Both rules are to some extent violated by isospin-non-conserving (mainly, Coulomb) 
interactions. {In the Z=N case, these violations appear as second order effects, 
while in mirror nuclei the effect is of first order. The difference is due to the 
interference between the {\em irregular} amplitude (symmetric with respect to the 
exchange of the two nuclei in the doublet) with the regular amplitude (which is 
isovector, antisymmetric with respect to the exchange).}

In the following, we discuss the relative importance of different possible
sources of asymmetry in mirror E1 transitions. As a simple example, we consider 
in particular those nuclei which can be described by the nuclear shell model
in a limited Hilbert space, containing a full major shell and the 
unique-parity intruder from the next major shell. Although the particle-hole 
excitations involving all states of the higher shell must be considered for 
a reliable description of the E1 transitions, we assume that the largest
part of the E1 amplitudes only involves the intruder orbital $j_I$ and, as a
consequence, only the largest-j orbital, $j_N=j_I-1$, of the lower major shell.
It is important to note that the inclusion of more orbitals in the calculation,
 briefly 
discussed in Appendix \ref{S:a3}, does not change substantially most of 
the results.

Actually, strong asymmetries in B(E1) values have been observed in 
several mirror transitions, e.g. pairs of mirror nuclei of the sd and
pf major shells \cite{oldmirror,letter}.
 The clearest examples, however, were found in light $N\simeq Z$ nuclei, 
such as $^{17}O$ and $^{17}F$. Such nuclei often exhibit large differences 
in the neutron and proton binding energies, and coupling to the continuum 
needs to be taken into account. The present discussion is limited instead to 
heavier mirror nuclei, in which the smaller binding energy of the proton is 
compensated by the larger coulomb barrier.

As a typical example (``benchmark'' in this work), we consider the mirror pair
 $^{67}$As - $^{67}$Se, 
{whose structure involves the pf shell plus the
$g_{9/2}$ intruder orbital. This doublet 
 has been investigated in a recent
 experiment at GAMMASPHERE \cite{letter}.  Two pairs of mirror transitions
with a sizable E1 component have been observed, connecting the lowest $9/2^+$
state to lower lying $7/2^-$ levels (Fig.~\ref{F:1}). The measured E1 strengths
and the absolute value of the corresponding E1 matrix elements are reported in
Table~\ref{T:1}. The $9/2^+$ state has presumably a rather pure $g_{9/2}$ 
character, while the daughter states have a complex structure and 
contain only a small component that can be reached by the E1 transition.
As a consequence, the observed values of B(E1) are very small.
}
{
All numerical results reported in the following will refer to this particular
pair of nuclei. The radial integrals have been obtained with 
single-particle wavefunctions in a Woods-Saxon potential with spin-orbit
interaction, as specified in \cite{BM}. These integrals change slowly
with the atomic number, and for the $f_{7/2}\to d_{5/2}$ transitions in the
middle of the $sd$ shell they would give results very close to those of the
$g_{9/2} \to f_{7/2}$ transitions in mass A=67.}

{
In Section \ref{S:1}, we derive the expression of the E1 
transition amplitude from the intruder state $a\ (J_a=j_I)$ to one
of the normal-parity states $b\ (J_b)$. No specific assumptions are made on the
structure of these states, apart from the fact that orbits of the higher 
major shell, different from the intruder, give negligible contribution.
}

In the following sections, we discuss the different processes that can lead 
to the presence
of an (induced) isoscalar
 E1 transition amplitude, in addition to the main isovector term.
In Section \ref{S:2} we consider the effect 
of higher order terms in the {nucleonic current, in addition to} those considered 
in the Siegert theorem,
which are increasingly important when the long-wavelength assumption fails.

In Section \ref{S:3} we discuss several simple effects related to the  
mixing of wavefunctions:
the Coulomb mixing between neighboring states (\ref{S:4.1}) and between
states of very similar structure, such as the analogue--antianalogue mixing
(\ref{S:4.2}).
None of the processes considered up to this point seems  able to justify
the observed asymmetry. We can conclude that the difference in the wavefunctions
of the two mirror nuclei involves many (weak) mixing with a large number of
states, possibly lying rather far in energy from the levels considered. There are 
two different approaches to consider this situation. {In the most direct treatment}
,  the residual interactions in the two mirror nuclei { are assumed from the start 
to be different and to include }the Coulomb interaction
(as well as other possible isospin violating terms). It is well known that 
most of the E1 strength is shifted to higher-lying {\em collective}
states, while the low-lying E1 transitions remain substantially hindered,
due to the destructive interference among the individual contributions.
If the residual interactions are not identical in the two mirror nuclei, the 
negative interference can amplify substantially these differences in the resulting 
B(E1). This mechanism is easy to understand, but even if a shell model calculation 
in this necessarily
huge Hilbert space were to become possible, the results would scarcely be}
 transparent with respect
to the nature of the processes involved. One could consider, however, the {\em same} 
problem from a different point of view. Namely, {let one suppose that a
zeroth-order calculation were} performed with isospin-conserving residual
interaction. As a next approximation, Coulomb interactions could be included
to evaluate, to first order, the mixing among {zeroth-order} states.
As in the former approach, one should expect that coherent contributions
from collective states play an significant role in producing the E1 asymmetry, 
as well as the concentration of the  E1 strength in the collective states 
has a role
in depleting the E1 strengths of low-lying transitions.
The advantage of this approach is that it can give semi-quantitative 
predictions on the B(E1) asymmetries, even without knowledge
of their absolute value. Furthermore, it would elucidate the principal process  
(or processes)
responsible for the largest part of the observed effects.

{A process of such kind, which could in principle account for the magnitude of the
observed effects} (namely, the coherent
contribution of states belonging to the giant isovector monopole resonance)
is discussed in detail in  section \ref{S:4.3}.

\begin{table}[t]
\caption{\label{T:1}Values of B(E1) for the transitions proceeding
from the lowest $9/2^+$ state in$^{67}$Se and $^{67}$As, as 
{
deduced from lifetimes and M2/E1 ratios, } determined
in \cite{letter}.
}
\vspace{0.4cm}
\begin{tabular}{cccc}
\hline\\[-3mm]
Nucleus & E$_\gamma$ & B(E1)
&$| (\frac{9}{2}^+ |\!| \mathcal{M}(E1) |\!| \frac{7}{2}^-) |$\\
& [keV] &[$e^2$ fm$^2$]& [e fm]\\[1mm]
\hline\\[-3mm]
$^{67}$As & $725$ & $1.4\pm 0.4 \times 10^{-6}$&$3.7\pm 0.5\times 10^{-3}$\\
$^{67}$Se & $717$ & $0.4\pm 0.4
 \times 10^{-6}$&$2.0\pm 2.0\times 10^{-3}$\\[1mm]
$^{67}$As & $319$ & $8.3\pm 2.5
 \times 10^{-6}$&$9.1\pm 1.3\times 10^{-3}$\\
$^{67}$Se & $303$ & $<1.4(9) \times 10^{-6}$&$< 3.7(11)\times 10^{-3}$\\
\hline\\
\end{tabular}
\end{table}

\section{\label{S:1}Isovector and isoscalar contributions}

{
In the following calculations, the E1 transition is assumed 
to take place  from  
an intruder single-particle orbital 
$j_I$ to a 
{
 normal-party orbital $j_N=j_I-1$ (or vice-versa)}.
The parent state $a$ will be the lowest intruder, with $J_a=j_I$
and parity $\bar{\pi}$. A possible daughter state $b$ must have 
$J_b=J_a\pm 1$ or $J_a$ and parity $\pi=-\bar{\pi}$. Its wave function can 
contain pairs inside intruder orbitals coupled to zero: in this case, the 
transition could proceed from a $j_N$ orbital present in $a$ to a $j_I$
orbital in $b$.}

 If we expand the 
wave functions of 
states $a$  and $b$
in terms of products of the one-body wave function times the core wave function
(with the proper fractional parentage coefficients) the only terms of the 
expansion
that contribute to the transition are those having a common core state
(of isospin $T_c = 0$ or 1) for both states $a$ and
$b$, and the single-particle orbit changing  from 
 {
$j_I$ to $j_N$} (with a core state $J_\mu^+$ of positive parity)
or vice-versa  (with a core state $J_\mu^-$ of negative parity):
\begin{eqnarray}
\label{E:1}
|a;J_a,M_a;1/2,T_3> &=&\\
&&\hspace*{-38mm}\phantom{+}\sum_\mu C_{fp}(a|j_I;\mu,J_\mu^+,T_c)
 [\phi(j_I) \otimes \Phi(\mu,J_\mu^+,T_c)]^{(J_a,1/2)}_{M_a,T_3} \nonumber \\
&&\hspace*{-43mm}\phantom{\times \Big[ }+ 
\sum_\mu C_{fp}(a|j_N;\mu,J_\mu^-,T_c)  
[\phi(j_N) \otimes \Phi(\mu,J_\mu^-,T_c)]^{(J_a,1/2)}_{M_a,T_3}\nonumber \\
&&\hspace*{-38mm} + ... \nonumber
\end{eqnarray}
\begin{eqnarray}
\label{E:2}
|b;J_b,M_b;1/2,T_3> &=&\\
&&\hspace*{-38mm}\phantom{+}\sum_\mu C_{fp}(b|j_N;\mu,J_\mu^+,T_c)
 [\phi(j_N) \otimes \Phi(\mu,J_\mu^+,T_c)]^{(J_b,1/2)}_{M_b,T_3}\nonumber\\
&&\hspace*{-43mm}\phantom{\times \Big[ }+ 
\sum_\mu C_{fp}(b|j_I;\mu,J_\mu^-,T_c)  
[\phi(j_I) \otimes \Phi(\mu,J_\mu^-,T_c)]^{(J_b,1/2)}_{M_b,T_3}\nonumber\\
&&\hspace*{-38mm} + ... \nonumber
\end{eqnarray}
Taking into account the relation, to be used  both in ordinary space and
in isospin space,
\begin{eqnarray}\label{E:3}
&&([j_1\otimes j_2]J |\!| U^{(K)}(1) |\!| [j_1^\prime\otimes j_2]J^\prime )
= (-1)^{j_1+j_2+J^\prime+K}\phantom{mm}\nonumber\\
&&\times \sqrt{(2J+1)(2J^\prime+1)}
\left\{ \begin{array}{c c c}
j_1&J&j_2\cr
J^\prime&j_1^\prime&K
\end{array}
\right\}
(j_1|\!| U^{(K)} |\!|j_1^\prime)\ ,
\end{eqnarray}
where $U^{(K)}$ is a tensor operator of rank $K$ acting only on the 
subspace ``1'',
we obtain for the reduced matrix element between states $a$ and $b$
\begin{eqnarray}\label{E:4}
&&\hspace*{-5mm}(b,J_b,T_b|\!|\!|\mathcal{M}_E^{(1K)}|\!|\!|a,J_a,T_a)
= \nonumber \\
&& \sum_{T_c} (-1)^{T_c+K+1}\widehat{T}_a\widehat{T}_b 
\ \left\{ \begin{array}{c c c}
1/2&T_b&T_c\cr
T_a&1/2&K
\end{array}
\right\}\phantom{m}\nonumber \\*
&&\hspace*{-5mm}\times\Big[ \sum_\mu (-1)^{J_a+j_N+J_\mu^++1}
\ \widehat{J}_a \widehat{J}_b
\left\{ \begin{array}{c c c}
j_N&J_b&J_\mu^+\cr
J_a&j_I&1
\end{array}
\right\}  \nonumber  \\*
&&\hspace*{-5mm}\phantom{m}\times n\  C_{fp}(a|j_I;\mu,J_\mu^+,T_c)
\ C_{fp}(b|j_N;\mu,J_\mu^+,T_c)\nonumber\\
&&\hspace*{-5mm}\phantom{m}\times 
(j_N|\!|\!|\mathcal{M}_E^{(1K)}|\!|\!|j_I)
\nonumber\\ 
&&\hspace*{-5mm}+\sum_\mu (-1)^{J_a+j_I+J_\mu^-+1}\ \widehat{J}_a \widehat{J}_b
\left\{ \begin{array}{c c c}
j_I&J_b&J_\mu^-\cr
J_a&j_N&1
\end{array}
\right\} \nonumber \\
&&\hspace*{-5mm}\phantom{m}\times n\ C_{fp}(a|j_N;\mu,J_\mu^-,T_c)\ C_{fp}
(b|j_I;\mu,J_\mu^-,T_c)\nonumber\\
&&\hspace*{-5mm}\phantom{m}\times 
(j_I|\!|\!|\mathcal{M}_E^{(1K)}|\!|\!|j_N)
\ \Big] \ ,
\end{eqnarray}
where  $n$ is the number of active nucleons, $\widehat{J}\equiv \sqrt{2J+1}$, 
and the triple bars indicate a reduced matrix element
with respect to ordinary space and to  isospin space (with 
$T_a=T_b=1/2$). 
The operator $\mathcal{M}_E^{(1K)}$ is now a tensor of rank 1 in the 
ordinary space and $K= 1$ or $0$ in isospin space. \\
Now, 
\begin{eqnarray}
\hspace*{-2.5mm}
(j_N|\!|\!|\mathcal{M}_E^{(1K)}|\!|\!|j_I)&=&\!(-1)^{j_I-j_N}
\overline{(j_I|\!|\!|\mathcal{M}_E^{(1K)}|\!|\!|j_N)}
\nonumber\\
&=&+\ (j_I|\!|\!|\mathcal{M}_E^{(1K)}|\!|\!|j_N)\phantom{i}.
\label{E:5}
\end{eqnarray}
Here, $j_I-j_N=1$, both single-particle states have isospin 1/2 and the 
E1 operator is odd under time reversal.
We obtain therefore for the reduced matrix element in ordinary space
\begin{eqnarray}\label{E:6}
&&(b,J_b;T_b,T_3|\!|\mathcal{M}_E^{(1K)}|\!|a,J_a;T_a,T_3)=
\nonumber \\
&=& (-1)^{1/2-T_3}
\left( \begin{array}{c c c}
T_b&K&T_a\cr
-T_3&0&T_3
\end{array}\right) (j_N|\!|\!|\mathcal{M}_E^{(1K)}|\!|\!|j_I)
\nonumber\\
&&\times  (-1)^K \sum_{T_c=0,1} \mathcal{A}(T_c)
\left\{ \begin{array}{c c c} 
1/2&T_b&T_c\cr
T_a&1/2&K
\end{array}\right\} \ ,
\end{eqnarray}

where $\mathcal{M}^{(11)}\equiv D_{IV}^{(1)} \vec{\tau}$ and
$\mathcal{M}^{(10)}\equiv D_{IS}^{(1)}$ are the isovector and isoscalar part
of the single-particle electric dipole operator:
\begin{eqnarray}\label{E:7}
(j_N|\!|\!|\mathcal{M}_E^{(10)}|\!|\!|j_I)&=&(j_N|\!|D^{(1)}_{IS}|
\!|j_I)(1/2|\!|1|\!|1/2)
\phantom{mm}\\
(j_N|\!|\!|\mathcal{M}_E^{(11)}|\!|\!|j_I)&=&
(j_N|\!|D^{(1)}_{IV}|\!|j_I)(1/2|\!|\vec{\tau}|\!|1/2)
\nonumber
\end{eqnarray}
and the core-isospin dependent coefficients $\mathcal{A}(T_c)$ ($T_c=0$ or $1$)
are
\begin{eqnarray}\label{E:8}
\mathcal{A}(T_c)&=&(-1)^{T_c}\ 2 \widehat{J}_a \widehat{J}_b\\
&\times& \Big[ \sum_\mu (-1)^{J_\mu^++J_a+j_N} \left\{ \begin{array}{c c c}
j_N&J_b&J_\mu^+\cr
J_a&j_I&1
\end{array}
\right\}  \nonumber \\ 
&\times& n C_{fp}(a|j_I;\mu,J_\mu^+,T_c)\times 
C_{fp}(b|j_N;\mu,J_\mu^+,T_c)\nonumber \\
&+&\sum_\mu (-1)^{J_\mu^-+J_b+J_I}
\left\{ \begin{array}{c c c}
j_I&J_a&J_\mu^-\cr
J_b&j_N&1
\end{array}
\right\} \nonumber \\
&\times& n  C_{fp}(a|j_N;\mu,J_\mu^-,T_c)\times 
C_{fp}(b|j_I;\mu,J_\mu^-,T_c)\Big]\ .
\nonumber 
\end{eqnarray}
With $T_a=T_b=1/2$, $T_c=0$ or $1$, by inserting the numerical 
 values of  the coefficients\footnote{
Namely, 
$(1/2|\!|1|\!|1/2)=\sqrt{2}$;\ $(1/2|\!|\vec{\tau}|\!|1/2)=\sqrt{6}$;
\ $\left( 
\begin{array}{ccc}
1/2&K&1/2\cr
-T_3&0&T_3
\end{array} \right)=(-1)^{1/2-T_3}/\sqrt{2}$ for $K=0$, and $=1/\sqrt{6}$
for $K=1$;\  $\left\{
\begin{array}{ccc}
1/2&K&T_c\cr
1/2&1/2&K
\end{array}\right\}=1/6$ for $K=T_c=1$ and $=(-1)^{T_c-1}/2$ for all other 
cases.} 
 one obtains
\begin{eqnarray}
\label{E:10}
&&(b,J_b;1/2,T_3|\!|\mathcal{M}_E^{(11)}|\!|a,J_a;1/2,T_3)\\
&&\phantom{mmm}=
\frac{(-1)^{1/2+T_3}}{6} \left[ \mathcal{A}(1) + 3\mathcal{A}(0)
\right]  (j_I|\!|D^{(1)}_{IV}|\!|j_N)\nonumber\\
\label{E:11}
&&(b,J_b;1/2,T_3|\!|\mathcal{M}_E^{(10)}|\!|a,J_a;1/2,T_3)\\
&&\phantom{mmm}=\frac{1}{2}
\left[ \mathcal{A}(1)-\mathcal{A}(0) \right]   
(j_I|\!|D^{(1)}_{IS}|\!|j_N)
\ .\nonumber
\end{eqnarray}
The leading isovector term in the single-particle operator is, in our case,
\begin{equation}
\hspace*{-2mm}
(j_I|\!| D^{(1)}_{IV}|\!|j_N)=
\frac{e}{2} <j_I|r|j_N> (j_I|\!| Y^{(1)}|\!|j_N)
\label{E:12}
\end{equation}
where $Y^{(1)}$ is the spherical harmonic for $\ell =1$.

The different forms of possible isoscalar contributions are discussed
 in the following sections.

\section{\label{S:2}Higher-order terms after Siegert}

It is well known that the usual expression of electric transition amplitudes,
deduced from the Siegert's theorem, is only valid in the long wavelength
limit. The complete expression for the electric transition amplitude,
taking also into account relativistic corrections, is given by Friar and
Fallieros \cite{friar}:
\begin{eqnarray}\label{E:s1}
T(E,LM)&=&
\frac{k^{L-1}}{(2L+1)!!}\nonumber\\
&& \hspace*{-9mm}\times \int {\rm d}r^3 \left[ i\sqrt{\frac{L+1}{L}}
\dot{\rho}(\vec{r}) r^L Y^{(L)}_M(\hat{r}) g_L(kr)
\right. \nonumber\\
&&\hspace*{4mm}\left.
+\frac{2k^2r}{L+2} \vec{\mu}(\vec{r})\cdot \vec{\mathcal{Y}}^{(L,1)L}_M
h_L(kr) 
\right] \ ,
\end{eqnarray}
where $\vec{\mathcal{Y}}^{(L,1)L}_M$ is the vector spherical harmonic and
\begin{eqnarray}\label{E:s2}
g_L(z)&\approx & 1 -\frac{Lz^2}{2(L+2)(2L+3)} + ...\\
h_L(z)&=&-\frac{L+2}{Lz}\frac{\rm d}{{\rm d}z}\left\{ z^{-2L}  
\frac{\rm d}{{\rm d}z} \left[ z^{2L+1} g_L(z)\right]\right\}
\approx 1 + ...\nonumber
\end{eqnarray}  
For our purposes, it will be sufficient to consider only the first
term after the Siegert limit, as given in the Eqs.~(\ref{E:s2}).
We will consider first the part of the integral (\ref{E:s1})
containing the time derivative of the charge density 
$\rho = \rho_0(\vec{r}) \exp(-ikct)$,
 approximating the nucleus to an ensemble of point-like nucleons:
\begin{eqnarray}\label{E:s3}
\dot{\rho}=-i kc \rho(\vec{r})\approx -i kc \sum \frac{1+ \tau_3(i)}{2}
e\ \delta(\vec{r}-\vec{r_i})\ .
\end{eqnarray}
The isovector part  of the E1 transition operator is, 
for a single-particle transition,
\begin{eqnarray}
D_{IV}(E1)\tau_3=\frac{\sqrt{2}}{3} \frac{e}{2} kc\ r Y^{(1)}(\hat{r})\tau_3
\ .
\end{eqnarray}
{To} first approximation, the
isoscalar part of the transition amplitude only comes from the 
second term of the series expansion  of $g_L(kr)$.
As we have to deal with a one-body operator, we can easily obtain
the amount of this correction with respect to the main (Siegert) term,
for each single-particle transition:
\begin{eqnarray}\label{E:s4}
\frac{\left< \ell_1 ,j_1 |\!| D^E_{IS}(E1) |\!| \ell_2, j_2 \right>
}{\left< \ell_1 ,j_1 |\!| D_{IV}(E1) |\!| \ell_2, j_2 \right>}
=-\frac{k^2}{30}\frac{ \left< \ell_1,j_1 | r^3 | \ell_2, j_2 \right>  }%
{ \left< \ell_1,j_1 | r | \ell_2,j_2 \right> } \ .
\end{eqnarray}
{
Now we can estimate the numerical value of this ratio for the case
of the $A=67$ doublet \cite{letter} chosen as a suitable benchmark,
and for  a $g_{9/2} \to f_{7/2}$ transition. With Woods-Saxon radial 
wavefunctions one obtains
}
\begin{equation}\label{E:s5}
\frac{\left <f_{7/2}|(r/R_0)^3|g_{9/2}\right>}{
\left<f_{7/2}|r/R_0|g_{9/2}\right>} =
 0.834.
\end{equation}
Assuming $R_0=1.27 A^{1/3}$fm $=5.158$fm:
\begin{eqnarray}\label{E:s6}
\frac{k^2}{30}\ \frac{\left< f_{7/2}|r^3|g_{9/2}\right>}{
\left< f_{7/2}|r|g_{9/2}\right>} &=& 
\frac{1}{30}\ (kR_0)^2 \frac{\left< f_{7/2}|(r/R_0)^3|g_{9/2}\right>}{
\left< f_{7/2}|r/R_0|g_{9/2}\right>}\nonumber\\
&=&1.90 \times 10^{-5}(E_\gamma[{\rm MeV}])^2\ .
\end{eqnarray}

The evaluation of the second part of Eq.~(\ref{E:s1})
can be easily performed if we substitute the continuous magnetic density
$\vec{\mu} (\vec{r})$ with that of an ensemble of point-like nucleons with 
spin:

\begin{eqnarray}\label{E:s7}
\vec{\mu}(\vec{r})&=&\mu_n \sum_i \left\{ \frac{1+\tau_3(i)}{2}
\left[\vec{\ell}_i + g_p\vec{s}_i \right]\right. \nonumber\\
&+&  \left. \frac{1-\tau_3(i)}{2}
 g_n\vec{s}_i \right\} \delta(\vec{r}-\vec{r}_i)\nonumber \\
&=&
\frac{\mu_n}{2}\sum \big\{ 
\left[ \vec{j}_i +(g_p-1+g_n)\vec{s}_i\right] \nonumber\\
&+&  \left[ \vec{j}_i +(g_p-1-g_n)\vec{s}_i\right] \tau_3(i) 
\big\} \ ,
\end{eqnarray}
where $\mu_n=e\hbar/(2M_p)$ is the nuclear magneton (and 
$M_p$ the proton mass).
On the basis of Eq.~(\ref{E:s6}), we can observe 
that the magnitude of this
term in comparison to the first term of Eq.~(\ref{E:s1})
is given by
\begin{eqnarray}\label{E:s9}
\frac{k^2 \mu_n}{kec}=\frac{\hbar k c}{2M_p c^2}\approx 
 0.53\ 10^{-3}E_\gamma [MeV] \ .
\end {eqnarray}
Here, we are only interested in the isoscalar part, where 
 the contribution of the term $\vec{s}$ is hindered due to the 
numerical factor $g_p-1+g_n\approx 0.76$. 
The evaluation of the matrix elements
of $\vec{s}\cdot \vec{\mathcal{Y}}^{(L,1)L}_M$ and 
$\vec{j}\cdot \vec{\mathcal{Y}}^{(L,1)L}_M$
is performed in detail in Appendix \ref{S:am}.
{
For our benchmark,
corresponding to a $g_{9/2}\to f_{7/2}$ single-particle transition,
}
one obtains
\begin{eqnarray}\label{E:s10}
\frac{(g_{9/2} |\!| D^M_{IS} |\!| f_{7/2})}{(g_{9/2} 
|\!| D_{IV} |\!| f_{7/2})}
&\approx&-\frac{2\sqrt{2}}{3}\ \frac{k\mu_n}{ec}
\left[\frac{(g_p-1+g_n)}{\sqrt{6}} -\frac{1}{\sqrt{2}}\right]
\nonumber\\
&\approx& 2\times 10^{-4}\ (E_\gamma[{\rm MeV}])
\end{eqnarray}
if we assume that the above description of the magnetic density is
approximately correct.

For $\gamma$-ray energies around 1 MeV, both correction terms
are far too small to justify the observed asymmetries in $A=67$.

\section{\label{S:3}The Coulomb mixing of wave functions}

If { one takes into account the level mixing} due to the Coulomb interaction $V_c$,
the wavefunction of a pure eigenstate $|a_0>$ of the charge--invariant
Hamiltonian is changed into a new
one, $|a^\prime>$. To first order, 
\begin{eqnarray}
\label{E:30}
|a^\prime> &=& |a_0> + \sum_k \frac{<a_k |V_c|a_0>}{E(a_0)-E(a_k)}|a_k> \ ,
\end{eqnarray}
where the sum is extended over all states $|a_k>$ having the same $J^\pi$ as
$|a_0>$, and which may or may not have the same isospin.
The E1 transition matrix element between the modified states 
$a^\prime$, $b^\prime$
is, again to first order,
\begin{eqnarray}
\label{E:31}
&&<b^\prime|\mathcal{M}(E1) |a^\prime> 
= <b_0|\mathcal{M}(E1) |a_0>\phantom{mmmmmm}\\
&&\phantom{mmm}+\sum \frac{<a_k |V_c|a_0>}{E(a_0)-E(a_k)}
<b_0|\mathcal{M}(E1) |a_k> \nonumber \\
&&\phantom{mmm}+\sum \frac{<b_0 |V_c|b_k>}{E(b_0)-E(b_k)}
<b_k|\mathcal{M}(E1) |a_0> \nonumber\\
&&\phantom{mmm}\equiv <b_0|\mathcal{M}(E1) |a_0> + 
<b_0|\widetilde{\mathcal{M}}(E1) |a_0> \ . \nonumber 
\end{eqnarray}
It was assumed, here, that 
the $\mathcal{M}(E1)$ operator is pure isovector.
The ensemble of the first-order corrections (shortly indicated as 
$<b_0|\widetilde{\mathcal{M}}(E1) |a_0>$) transforms as an even tensor in
isospin space. In the $T=1/2$ or $T=0$ subspaces, it can be considered as an {\em 
induced isoscalar} amplitude.

 If $T_3=0$ and the {\em unperturbed} states
$a_0,\ b_0$ have the same isospin, the first term of the sum (\ref{E:31})
vanishes and only
the induced part contributes.
Instead, if $T_3=\pm 1/2$, the first term is the leading one and the other two
are only first-order corrections.

The Coulomb potential can be written as the sum of
 an isoscalar, an isovector and a rank-2 isotensor
term:
\begin{eqnarray}
\label{E:32}
V_c&=&\frac{1}{2}\sum_i \sum_{j\neq i} \frac{e^2}{r_{ij}}
\frac{1+\tau_3(i)}{2} \frac{1+\tau_3(j)}{2}\\
&=&\frac{1}{8}\sum_i \sum_{j\neq i} \frac{e^2}{r_{ij}}
\left[ 1+\frac{1}{3}(\bm{\tau} (i)\cdot \bm{\tau}(j)\ )\right]\nonumber\\
&+&\sum_i \frac{e}{2}\tau_3(i) \sum_{j\neq i}  \frac{e}{2} \frac{1}{r_{ij}} 
\nonumber\\
&+&\frac{1}{8}\sum_i \sum_{j\neq i} \frac{e^2}{r_{ij}}\Big[ 
\tau_3(i)\tau_3(j)-
\frac{1}{3}\big(\ \bm{\tau} (i)\cdot \bm{\tau}(j)\ \big)\Big] \ .\nonumber
\end{eqnarray}
The isoscalar part can be included in the charge-invariant Hamiltonian.
The matrix elements of the isotensor term vanish in the $T=1/2$ subspace.
They could contribute to the mixing with a $T=3/2$ state but would 
 produce, in any case, equal effects in two mirror nuclei.

Therefore, any difference between mirror nuclei
has to be attributed to the mixing induced by the isovector term
$V_c^{(1)}$
\begin{eqnarray}
\label{E:33}
V_c^{(1)}&=&\sum_i \frac{e}{2}\tau_3(i) \sum_{j\neq i}  \frac{e}{2} 
\frac{1}{r_{ij}} \ , 
\end{eqnarray}
where $V_c^{(1)}$ is, obviously, a two-body operator. 
It is possible, however, 
to approximate its matrix elements with those of a suitable one-body 
operator (see~\cite{BM}, Eq. 2-104).
Actually, the second sum in Eq.~(\ref{E:33}) corresponds to the Coulomb 
potential of a system of $A-1$ point-like charges $e/2$ associated to all 
nucleons $j$ different from the nucleon $i$, and we can approximate it with the
electrostatic potential of an uniformly-charged sphere of radius $R$,
{\it i.e.}, for $r<R$
\begin{equation}
\varphi_c(r)\equiv \frac{e(A-1)}{2R}\ f_c(r/R)\approx 
\frac{e(A-1)}{2R}\ \frac{3R^2-r^2}{2R^2} \ .
\label{E:34}
\end{equation}
(slightly different forms of the function $f_c$ will be considered in
the following). With these approximations,
\begin{eqnarray}
\label{E:35}
V_c^{(1)}&\approx&\sum_i \frac{e}{2}\tau_3(i) \varphi_c(r_i)\\\
 &=&  e T_3 \varphi_c(0) - \sum_i 
\frac{e}{2}\tau_3(i)[\varphi_c(0)-\varphi_c(r)]\nonumber \\
&\equiv& e T_3 \varphi_c(0)+\widetilde{V}_c^{(1)}\nonumber
\end{eqnarray}
and, for the potential $\varphi_c$ of an uniformly charged sphere,
\begin{eqnarray}
\widetilde{V}_c^{(1)}=-\frac{e(A-1)}{R^3} \sum_i 
\frac{e r_i^2 \tau_3(i)}{8}  \ . \phantom{mm} 
\label{E:35b}
\end{eqnarray}
The first term of Eq. (\ref{E:35}) 
is diagonal and does not contribute to the mixing. The 
second term is proportional to the isovector monopole operator 
\begin{eqnarray}\label{E:36}
\mathcal{M}^{(1)}(E0)=\sum_i  \frac{e\ r_i^2 \tau_3(i)}{2} \ .
\end{eqnarray}
This result will be exploited again in Section \ref{S:4.3}.

Actually, the use of a constant charge density inside a sphere to evaluate the 
electrostatic potential $\varphi_c$ is somewhat inconsistent with the
Woods-Saxon distribution of matter density assumed to calculate the
radial wavefunctions. Moreover, the tails of these wavefunctions extend
outside the nuclear radius, in a region where $\varphi_c$ would 
decrease as $1/r$. Calculations of the electrostatic potential for a
Woods-Saxon density of charge are given in  Appendix \ref{S:a.2}. 
For small values of $r$ --  i.e., as long as the charge density
of the Woods-Saxon distribution is substantially constant and equal
to that of the sphere -- the values of $\widetilde{V}_c^{(1)}$ are equal in 
the two cases, and the differences in the calculated integral are
always  rather small. To obtain the same charge density at the centre, the
radius $R$ of the uniformly charged sphere { must take a slightly different value} 
from the parameter $R_0$ of the Woods-Saxon distribution.  Adopting for the 
Woods-Saxon parameters the values 
suggested by Bohr and Mottelson \cite{BM},
$R_0=1.27 A^{1/3}$ fm, $a=0.67$ fm, for $A=67$ one obtains
$R_0=5.158$~fm and $R=5.430$~fm.

The matrix elements of $\widetilde{V}_c^{(1)}$ are in any case 
very small.
To produce a sizable mixing of states, it is necessary that the effect
be amplified due to some particular conditions. This can happen, in
particular, (i) when two levels with equal $J^\pi$ are very close in
energy or (ii) have very similar  wavefunctions, or (iii) when many
different levels  contribute {\em coherently} to the mixing.
We will consider these three cases in the following subsections.

\subsection{\label{S:4.1}Close-lying states }
The simplest possible case is the mixing of two states which lie close in 
energy.
As an example, we can consider the E1 decay of a given state $a$  
(of spin $J_a$) towards two states
$b_1,\ b_2$ of equal angular momentum $J_b$, and rather close in energy.
In this case, taking into account only the Coulomb mixing between $b_1$ and 
$b_2$
(and neglecting small isoscalar terms in the E1
operator) we obtain up to  first order
\begin{eqnarray}\label{E:37}
&&(b_1^\prime,J_b|\!|\mathcal{M}(E1)|\!|a,J_a)=\\
&&=\!(b_1,J_b\!|\mathcal{M}(E1)|\!|a,J_a)
+\alpha T_3 (b_2,J_b|\!|\mathcal{M}(E1)|\!|a,J_a)\nonumber\\
&&(b_2^\prime,J_b|\!|\mathcal{M}(E1)|\!|a,J_a\!)=\\
&&=\!(b_2,J_b|\!|\mathcal{M}(E1)|\!|a,J_a)
-\alpha T_3 (b_1,J_b|\!|\mathcal{M}(E1)|\!|a,J_a)\nonumber\phantom{n}
\nonumber
\end{eqnarray}
with
\begin{equation}
\alpha T_3 = <b_1 | \widetilde{V}_c^{(1)} |b_2 > /\ [E(b_1)-E(b_2)]
\label{E:38}
\end{equation}
In fact, as a consequence of the Wigner--Eckart theorem, the matrix element of
$\widetilde{V}_c^{(1)}$ must be proportional to that of $T_3$.

The reduced transition probabilities become, up to first order
\begin{eqnarray}\label{E:39}
&&B(E1; a\to b^\prime_1)= \frac{1}{2J+1} 
\Big[(b_1,J^\prime|\!|\mathcal{M}(E1)|\!|a,J)^2\\
&&+2\alpha T_3 (b_1,J_b|\!|\mathcal{M}(E1)|\!|a,J_a)(b_2,J_b|\!|
\mathcal{M}(E1)|\!|a,J_a)\Big] \nonumber\\
\label{E:40}
&&B(E1; a \to b^\prime_2) = \frac{1}{2J_a+1}
\Big[ (b_2,J_b|\!|\mathcal{M}(E1)|\!|a,J_a)^2\\
&&-2\alpha T_3 (b_1,J_b|\!|\mathcal{M}(E1)|\!|a,J_a)(b_1,J_b\!|
\mathcal{M}(E1)|\!|a,J_a)\Big]\nonumber.
\end{eqnarray}
Hence, the sum of the two reduced strengths,
\begin{eqnarray}\label{E:41}
&&B(E1; a\to b^\prime_1) + B(E1; a\to b^\prime_2) =\\
&&\frac{1}{2J_a+1} \Big[ (b_1,J_b|\!|\mathcal{M}(E1)|\!|a,J_a)^2
+(b_2,J_b|\!|\mathcal{M}(E1)|\!|a,J_a)^2\ \Big] \nonumber
\end{eqnarray}
is independent of $T_3$ and consequently identical in the two mirror nuclei.
If one of the two unperturbed transition strengths (either for $a\to b_1$ 
or $a\to b_2$) is much smaller than the other, a large percentage difference
between mirror values can be found, but only for the weaker
transition.

\subsection{\label{S:4.2}Analogue -- antianalogue mixing}
A second interesting case concerns the mixing between two very similar
wavefunctions, as for a pair of {\em analogue -- antianalogue states}
(this would be a very favourable case of the mixing of $T=1/2$ and
$T=3/2$ states discussed in \cite{patt08}). 
Let us consider, as a simple example, the state obtained with the coupling
of a $j_I=9/2$ nucleon to the lowest state $\phi_0$ ($J^\pi=0^+,\ T=1$)
of the isospin triplet $A=66$. Isospin $3/2$ states are obtained in the two
$|T_3|=3/2$ nuclei. In the $|T_3|=1/2$ nuclei $^{67}$As, $^{67}$Se
two independent wavefunctions will
result from the coupling, and two pure isospin states
can be constructed by proper linear combinations: a $T=3/2$ 
state $|a_3>$,
which is the {\em isospin analogue} of those in the $|T_3|=3/2$ nuclei,
and a  $T=1/2$ state $|a_1>$, sometime referred to as the
{\em anti-analogue} of them. Here we will give the results for the $T_3=+1/2$
nucleus (from which, those for $T_3=-1/2$ can be easily deduced by means 
of the  Wigner-Eckart theorem):  
\begin{eqnarray}\label{E:42}
\left|a_3\right> &=&\big| [\phi_j(t=1/2)\otimes \Phi_0(T_c=1)]j,T=3/2\big>\\
&=&c_1 \big|\phi_\pi(g_{9/2})\ \Phi_0(T_3=0)\big>\nonumber \\
&+&c_2 \big| \phi_\nu(g_{9/2})\ \Phi_0(T_3=1)\big>\nonumber
\end{eqnarray}
\begin{eqnarray}
\label{E:antia}
\left|a_1\right> &=&\big|[\phi_j(t=1/2)\otimes \Phi_0(T_c=1)]j,T=1/2\big>\\
&=&c_2 \big| \phi_\pi(g_{9/2})\ \Phi_0(T_3=0)\big>\nonumber \\
&-&c_1 \big| \phi_\nu(g_{9/2})\ \Phi_0(T_3=1)\big>\nonumber
\end{eqnarray}
where, for $T_3=+1/2$,
\begin{eqnarray}
c_1&=&(1/2,1/2,1,0\ |\ 3/2,1/2)\nonumber\\
&=&-(1/2,-1/2,1,1\ |\ 1/2,1/2)
=\sqrt{2/3}
\end{eqnarray}
and
\begin{eqnarray}
c_2&=&(1/2,-1/2,1,1\ |\ 3/2,1/2)\nonumber\\
&=&(1/2,1/2,1,0\ |\ 1/2,1/2)
=\sqrt{1/3}.
\end{eqnarray}
We now use Eqs.~(\ref{E:35},\ref{E:35b}) 
to approximate the non diagonal part 
of the  isovector Coulomb interaction  $V_C^{(1)}$
with a one-body operator  $\widetilde{V}_C^{(1)}$, whose matrix element 
between analogue and antianalogue states is, for $T_3=+1/2$,
\begin{eqnarray}\label{E:43}
\left< a_3 | \widetilde{V}_C^{(1)} |a_1 \right> &=&
\!c_1c_2
\left<\phi_\pi \Phi_0(1,0)|\widetilde{V}_C^{(1)}|
\phi_\pi \Phi_0(1,0) \right> \nonumber \\
&-&\!c_2c_1 \left<\phi_\nu \Phi_0(1,1)|\widetilde{V}^{(1)}_C|
\phi_\nu \Phi_0(1,1)\right> \nonumber\\
&=&\!c_1c_2 \left[ \left< \phi|\widetilde{V}^{(1)}_C(\pi)|\phi \right>
- \left< \phi|\widetilde{V}^{(1)}_C(\mu)|\phi \right> \right.\nonumber \\ 
&-& \left. \left<\Phi_0(1,1)|\widetilde{V}^{(1)}_C|
\Phi_0(1,1)\right> \right]\ .
\end{eqnarray}
The diagonal matrix element of the {\em isovector} operator 
$\widetilde{V}^{(1)}_C$
over the core state  $T=1,T_3=0$ is zero.

Starting from Eq. (\ref{E:43}) and assuming an energy spacing
$E(a_3)-E(a_1)=\Delta E$,
 we can now estimate at least the order 
of magnitude of the mixing coefficient. For $T_3=+1/2$,
\begin{eqnarray}\label{E:44}
\alpha&=&\frac{\Big< a_3 \Big| \widetilde{V}^{(1)}_C 
\Big|a_1 \Big>}{(-\Delta E)}
=c_1c_2  \frac{(A-1)e^2}{8R\ \Delta E} \Big[ 2\Big< g_{9/2}
 | \frac{r^2}{R^2} | g_{9/2}
 \Big> \nonumber\\
&-& \Big< \Phi_0(1,1)\Big|\sum_i \tau_3(i)\frac{r_i^2}{R^2} \Big|   
\Phi_0(1,1)\Big>  \Big] 
\end{eqnarray}
In the second term, the contributions of a proton and
of a neutron {\em in the same orbit} cancel one another, due
to the opposite eigenvalue of $\tau_3$. There are, however, two
excess protons in the $T_3=1$ core state. If all the radial wavefunctions 
of active nucleons in the core were equivalent to that of the $j_I$ orbit,
the second term in the sum of Eq.~(\ref{E:44}) would exactly cancel the first 
one. We can expect, therefore, a resulting matrix element
substantially smaller than the first term alone, due to the
effect of the core term. However, the 
expectation value of $r^2/R^2$ in the $j_I=g_{9/2}$ orbit is certainly
larger than those for the lower orbits in the core. 
For $A=67$, and with Woods-Saxon wavefunctions,
the radial integral of $(r/R)^2$ in the $0g_{9/2}$ orbit is 
0.7495, 
while in the normal-party orbits  $0f_{7/2}$.
$0f_{5/2}$, $1p_{3/2}$ and $1p_{1/2}$ is, respectively,
0.6251, 0.5922, 0.6251, and 0.6359.
In Eq.~(\ref{E:44}), we will use the  
average of these values,
  $<r^2/R^2>=0.6119$, and
the above estimate of the matrix element in the $g_{9/2}$ orbit,
to evaluate an order of magnitude for the analogue-antianalogue 
mixing\footnote{See Table \ref{tab:aa1} in the 
Appendix \ref{S:a.2}.
With a Woods-Saxon charge distribution, the
estimate does not change more than a few percent}. 
Numerically, with $c_1c_2=\sqrt{2}/3$, $A=67$,
$R=4.43$fm
and assuming $\Delta E\approx 4$~MeV as in $^{59}$Cu~\cite{maripuu72},
we obtain 
$\alpha \approx 0.071$. 
As the matrix element of the isovector interaction $\widetilde{V}^{(1)}_C$ 
between a state of isospin $3/2$ and a state of isospin $1/2$ is 
\begin{eqnarray}
\label{E:WE}
\left< 3/2, T_3| \widetilde{V}^{(1)}_C |1/2, T_3\right>
&=&(-1)^{3/2-T_3}\left(
\begin{array}{ccc}
3/2&1&1/2\cr
-T_3&0&T_3
\end{array}
\right)\nonumber\\
&\times& (3/2|\!|\widetilde{V}^{(1)}_C|\!|1/2)
\phantom{mm}
\end {eqnarray}
the value of $\alpha$ has equal sign in both nuclei of the doublet.

The E1 transition matrix element from the state $|a_1^\prime>$ to a given 
state $|b>$ will be, at the first order,
\begin{eqnarray}\label{E:45}
\left< b|\!|\mathcal{M}(E1)|\!| a_1^\prime  \right>
=\left< b|\!|\mathcal{M}(E1)|\!| a_1  \right>
+\alpha \left< b|\!|\mathcal{M}(E1)|\!| a_3  \right>.\phantom{mm}
\end{eqnarray}
We assume, for sake of simplicity, that the state $b$ has pure isospin 1/2. 
If, as we have supposed, the E1 transition proceeds from a
$j_I=9/2$ to a $j_N=7/2$ single-particle state, we can use 
for the state $|b>$ a fractional parentage expansion
in the style of the first line of Eq.~(\ref{E:2}). But
only the terms corresponding to the coupling of a nucleon in the state
$j_N=7/2$ to the core configuration $\Phi_0$ with $J=0,T=1$ 
can be reached by the E1 transition. We can write the
(presumably small) part of the wavefunction of the state $b$ which is 
relevant for the E1 transition in the form of Eq.(\ref{E:antia}).
\begin{eqnarray}
\left|b\right> &=&\left|[\phi_j(t=1/2)\otimes \Phi_0(T_c=1)]j,T=1/2\right>+ ...
\nonumber\\
&=&c_2 \left|\phi_\pi(f_{7/2})\  \Phi_0(T_3=0)\right> \\
&-& c_1 \left| \phi_\mu(f_{7/2})\ \Phi_0(T_3=1)\right>+ ...\nonumber
\end{eqnarray}
Taking into account the effective charges for E1 transition, 
$\epsilon_\pi=1/2$ and $\epsilon_\nu=-1/2$,
from Eq. (\ref{E:45}) we obtain
\begin{eqnarray}\label{E:anti2}
\left( b |\!|\mathcal{M}(E1)|\!|  a_1^\prime \right)
&=&\left[(|c_2|^2 \epsilon_\pi +|c_1|^2 \epsilon_\nu) 
+\alpha c_1c_2 ( \epsilon_\pi
-\epsilon_\nu) \right]\nonumber \\
&\times&
 ( f_{7/2}|\!| e r Y^{(1)}|\!| g_{9/2} )\nonumber\\
&&\hspace*{-12mm}=\frac{-1+2\sqrt{2}\ \alpha }{6} 
 ( f_{7/2}|\!| e r Y^{(1)}|\!| g_{9/2} ) \nonumber  
\end{eqnarray}
For $T_3=\pm 1/2$, using Eq.(\ref{E:WE}) we obtain
the numerical coefficient $(\mp 1+2\sqrt{2}\ \alpha)/6$.
In conclusion, the E1 strength in the two mirror transitions is proportional
to $(\mp 1+ 2\sqrt{2} \alpha)^2$.
The mirror asymmetry in the E1 strength is therefore, approximately,
\begin{equation}
\frac{B(E1,As)-B(E1,Se)}{B(E1,As)+B(E1,Se)}=\frac{4
\sqrt{2}\ \alpha}{1+8\alpha^2} \approx 0.386
\label{E:48}
\end{equation}
and the ratio  $B(E1,As)/B(E1,Se) \approx 2.26$.
We note, however, that such a large asymmetry has been obtained for
a pure configuration of the analogue and antianalogue states, while 
the antianalogue strength is usually spread over a number of final 
states~\cite{sherr65}, a situation which will strongly reduce the
mirror asymmetry in the E1 strength.
A detailed shell-model
investigation would possibly elucidate the role of the 
analogue-antianalogue mixing in the E1 asymmetry between mirror
nuclei, as the analogue and the antianalogue states can be described
in the same shell-model space.
  
\subsection{\label{S:4.3}Coherent enhancement of induced isoscalar E1}
The Coulomb mixing discussed in the previous subsections involves
states belonging to the same set of shell-model orbits necessary
for the (unperturbed) parent and daughter state of the E1 transitions
(presumably limited to two major shells).
However, it is well known that a comparatively large contribution to 
the isospin mixing comes from states outside this model space, as those
belonging to the giant isovector monopole resonance~\cite{colo95}.
Obviously,
the mixing with any of these higher-lying states, induced
by the isovector part of the Coulomb interaction, is
expected to be very small. The combined effect of many
higher-lying states on the E1 transition amplitude can
however become appreciable if their individual contributions
combine coherently. We shall see how this can be the case.

We have seen (Eqs.~\ref{E:35},\ref{E:35b})
 that the non-diagonal isovector part of the Coulomb 
interaction  $V_c$ can be approximated
with a one-body operator $\widetilde{V}_C^{(1)}$, 
having the same form of the isovector monopole
operator $\mathcal{M}^{(1)}(E0)$.
Therefore, it is a sensible approximation~\cite{bini,bizzeti}
to consider in the ensemble of states
$a_k$, $b_k$ (with $k\neq 0$)   of Eqs.~(\ref{E:30},\ref{E:31})
only those of the isovector monopole resonances built over $a_0$
and $b_0$, and to use the mean excitation energy $\Delta E_a$
(or $\Delta E_b$) of the giant resonance over the state $a_0$ (or $b_0$)
in the place of those of individual states.
In this case, Eq.~(\ref{E:31}) becomes
\begin{eqnarray}\label{E:49}
&&\hspace*{-8mm}<b^\prime|\mathcal{M}(E1) |a^\prime> =\\
&&= <b_0|\mathcal{M}(E1) |a_0>+
<b_0|\widetilde{\mathcal{M}}(E1) |a_0> \ ,\nonumber
\end{eqnarray}
where
\begin{eqnarray}\label{E:50}
&&\hspace*{-8mm}<b_0|\widetilde{\mathcal{M}}(E1) |a_0>\\
&\approx &\frac{-1}{\Delta E_a}\sum <b_0|\mathcal{M}(E1) |a_k><a_k 
|V_c|a_0> \nonumber\\
&+&\frac{-1}{\Delta E_b}\sum <b_0 |V_c|b_k><b_k|\mathcal{M}(E1) 
|a_0>\ . \nonumber
\end{eqnarray}
We are only interested in the isoscalar part of $\mathcal{M}(E1)$, which 
results from the isovector part $V_c^{(1)}$ of the Coulomb
interaction. Approximating  the non diagonal part of $V_c^{(1)}$ 
with the one-body potential of Eq. (\ref{E:35b}),
the {\em closure approximation} gives
\begin{eqnarray}\label{E:51}
&&\hspace*{-8mm}\sum <b_0 |\widetilde{V}^{(1)}_c|b_k><b_k|\mathcal{M}(E1) |a_0>
\nonumber\\
&\approx&<b_0 |\widetilde{V}^{(1)}_c\ \mathcal{M}(E1) |a_0>\\
&&\hspace*{-8mm}\sum <b_0|\mathcal{M}(E1) |a_k>
<a_k |\widetilde{V}^{(1)}_c|a_0>
\nonumber \\
&\approx&<b_0 |\mathcal{M}(E1)\ \widetilde{V}^{(1)}_c|a_0>\nonumber \ .
\end{eqnarray}
and therefore (as $\mathcal{M}(E1)$ and $ \widetilde{V}^{(1)}_c$ commute)
\begin{eqnarray}\label{E:52}
<b_0|\widetilde{\mathcal{M}}^{(0)}(E1) |a_0> 
&\approx&  2 \frac{<b_0 |\mathcal{M}(E1) \widetilde{V}_c^{(1)} 
|a_0>}{(-\Delta E_0)}
\phantom{mmm}
\end{eqnarray}
where we have assumed  $\Delta E_a \approx \Delta E_b \Rightarrow \Delta E_0$. 
\begin{eqnarray}
\label{E:53}
&&\hspace{-4mm}<b_0|\widetilde{\mathcal{M}}^{(0)}(E1) V_c^{(1)}|a_0>\approx  
\ \frac{2}{(-\Delta E_0)}\ \times\nonumber\\
&&\hspace{-4mm}
\Big< b_0 \Big| \sum_i \frac{e}{2}  r_i Y^{(1)}(\hat{r}_i)  \tau_3(i) 
\sum_j \frac{e}{2} [\varphi_c(r_j)-\varphi_c(0)]\tau_3(j) 
\Big| a_0 \Big> \phantom{n}\nonumber\\
&\equiv&
\ \Big< b_0 \Big|
\widetilde{\mathcal{M}}^{(0)}_{1-b} + \widetilde{\mathcal{M}}^{(0)}_{2-b}
\Big| a_0 \Big>
\end{eqnarray}
where  $\widetilde{\mathcal{M}}^{(0)}_{1-b}$ is the one-body operator resulting
from the term with $j=i$ in the second sum, and 
$\widetilde{\mathcal{M}}^{(0)}_{2-b}$
is a two-body operator resulting from all other terms.
As $\tau_3^2=1$, the first term is
\begin{eqnarray}
\label{E:54a}
\widetilde{\mathcal{M}}^{(0)}_{1-b}=\frac{1}{\Delta E_0}
\sum_i\ e[\varphi_c(0)-\varphi_c(r_i)]
\ \frac{e}{2} r_i Y^{(1)}(\hat{r}_i)
\end{eqnarray}
With the expression of $\varphi_c$ corresponding to the uniformly 
charged sphere, given in Eq. (\ref{E:34}) (and extrapolated also for $r>R$) 
one  obtains for  the one-body operator
\begin{eqnarray} 
\label{E:34b}
\widetilde{\mathcal{M}}^{(0)}_{1-b}\equiv C
\ \sum_i \frac{r_i^3}{R^2} \frac{e}{2}\ Y^{(1)}(\hat{r}_i)
\end{eqnarray}
which has the same structure 
as the one  coming from the second-order term
in the series expansion of Eq.~(\ref{E:s1}), with a different coefficient,
\begin{equation}
\label{E:34c}
C=+(A-1)e^2/\ ( 8 R \Delta E_0).
\end{equation}
An alternative calculation using a Woods-Saxon charge distribution is
reported in the Appendix \ref{S:a.2}.

As the one-body operator (\ref{E:52}) is isoscalar, its matrix elements 
can be expressed in the form anticipated in Eq.~(\ref{E:11}):
\begin{eqnarray}
\label{E:54}
&&\hspace*{-1mm}\Big(b,J_b;\frac{1}{2},T_3\Big|\!\Big|\sum_i \frac{e}{2}
\ \frac{r_i^2}{R^3} \ r_i
Y^{(1)}(\hat{r}_i)\Big|\!\Big|a,J_a;\frac{1}{2},T_3 \Big)\nonumber\\
&& \hspace*{10mm}= \frac{1}{2}\left[ \mathcal{A}(1)-\mathcal{A}(0) \right]  
\frac{e}{2}\ \Big( j_N\Big|\!\Big| r\  Y^{(1)}(\hat{r})
\Big|\!\Big|j_I\Big)\nonumber \\
&& \hspace*{15mm}\times
\frac{\left< j_N|(r/R)^3|j_I \right> }{\left< j_N|(r/R)|j_I \right> }
\end{eqnarray}
Again, we can evaluate the numerical results for our benchmark doublet.
For $A=67$, we assume $R = 5.30$ fm. The energy difference is
$\Delta E_0\approx 20$~MeV in $^{60}$Ni (according to \cite{nakayama}). As 
$\Delta E_0$ is expected to scale as $A^{-1/3}$~\cite{colo95}, we assume 
$\Delta E_0\approx 19.3$~MeV for $A=67$. With these assumptions,
the numerical value of the adimensional coefficient $C$ in
Eq. (\ref{E:34c}) is  $C=0.116$. 
For the ratio of radial integrals (last factor of Eq.~(\ref{E:54})),
with the radial wavefunctions corresponding to the Woods-Saxon potential
one obtains
$\left< g_{9/2}| (r/R)^3 | f_{7/2}\right> 
/ \left< g_{9/2}|r/R| f_{7/2}\right> = 0.752$.
 
It remains to consider the two-body term (second term of Eq.~(\ref{E:53})).
Again, we can use the fractional parentage expansion of Eqs.~(\ref{E:1},
\ref{E:2}).
Here, however, the tensor operator is the product of two factors: a vector 
isovector
term acting on the single-particle state and a scalar isovector one acting on
 the core state. The product $\tau_3(i) \tau_3(j)$ contains an isoscalar and
an isotensor part:
\begin{eqnarray}\label{E:56}
\tau_3(i) \tau_3(j)&=&
\Big[ \tau_3(i) \tau_3(j)-\ \frac{1}{3}\ \big(\vec{\tau}(i)\cdot 
\vec{\tau}(j)\big)
\Big] \nonumber\\
&+&\frac{1}{3}\ \big(\vec{\tau}(i)\cdot \vec{\tau}(j)\big)
\end{eqnarray}
but only the isoscalar is effective if the states $a_0$ and $b_0$ have $T=1/2$.
To evaluate the reduced matrix element for the isoscalar part
of the two-body operator
\begin{eqnarray}
\label{E:57}
\widetilde{\mathcal{M}}^{(0)}_{2-b}
&=&\frac{e^2}{6\Delta E_0} \sum_i 
r_i Y^{(1)}(\hat{r}_i)\\ 
&\times&\Big( \vec{\tau}(i) \cdot \sum_{j\neq i} 
\vec{\tau}(j) [\varphi(0)-\varphi_c(r_j)] \Big)\phantom{mmm}\nonumber
\end{eqnarray}
we can use the standard relations of tensor algebra 
for the matrix elements of tensor products
to obtain the reduced matrix element (in ordinary space)\footnote{ 
In fact:
$
\ \left< 1/2,T_3;1/2,T_3 | (\vec{\tau}(i)\cdot \vec{\tau}(j)
| 1/2,T_c^\prime;1/2,T_3  \right>
=(-1)^{T_c+1}
 \left\{
\begin{array}{ccc}
1/2&T_c&1/2\cr
T_c^\prime&1/2&1
\end{array}\right\}
(1/2|\!|\vec{\tau}(i)|\!|1/2) (T_c|\!|\vec{\tau}(i)|\!|T_c^\prime)
$;\ and\\
 $
(j_N,J_\mu,J_b|\!|r_iY^{(1)}(i)
 (r_j/R)^2)|\!|j_I.J_\mu,J_a)=\widehat{J}_a \widehat{J}_b \\
\times
 \left\{
\begin{array}{ccc}
j_N&J_\mu&J_b\cr
j_I&J_\mu&J_a\cr
1&0&1
\end{array}
\right\} 
(j_N|\!|r_iY^{(1)}(i)|\!|j_I)\ (J_\mu|\!|(r_j/R)^2 |\!|J_\mu)
$
}:
\begin{eqnarray}\label{E:60}
&&\Big( b_0;J_b;1/2,T_3 \Big|\!\Big| 
\widetilde{\mathcal{M}}^{(0)}_{2-b}
\Big|\!\Big| a_0;J_a,1/2,T_3 \Big)\nonumber\\
&=&\frac{e}{2}\big( j_N |\!| r Y^{(1)}(\hat{r})  |\!| j_I \big)
\nonumber\\
&\times& \ \frac{C}{\sqrt{3}}\ \sum_{T_c,T_c^\prime} (-1)^{T_c+1}
\left\{ \begin{array}{ccc}
1/2&T_c&1/2\cr
T_c^\prime&1/2&1
\end{array} \right\}\\
&\times& \Bigg[ \sum_{\mu,\mu^\prime} (-1)^{j_N+J_a+J_\mu^++1}
\frac{\widehat{J}_a\widehat{J}_b}
{\widehat{J}^+_\mu}
\left\{ \begin{array}{ccc}
J_b&j_N&J_\mu^+\cr
j_I&J_a&1
\end{array} \right\} \times
\nonumber\\
&&\phantom{\times}\ C_{fp}(a|j_I;\mu,J^+_\mu,T_c)
\ C_{fp}(b|j_N;\mu^\prime,J^+_\mu,T_c^\prime)  \nonumber\\
&&\phantom{\times}\times
\big(\mu,J^+_\mu,T_c |\!|\!|\sum \frac{r_j^2}{R^2}
 \vec{\tau} (j) |\!|\!|\mu^\prime ,J^+_\mu,T_c^\prime \big)\nonumber\\
&&+ \sum_{\mu,\mu^\prime}  (-1)^{j_I+J_a+J_\mu^-+1}
\frac{\widehat{J}_a\widehat{J}_b}{\widehat{J}^-_\mu} 
\left\{ \begin{array}{ccc}
J_b&j_I&J^-_\mu\cr
j_N&J_a&1
\end{array} \right\} \times
\nonumber\\
&&\phantom{\times}\ C_{fp}(a|j_N;\mu,J^-_\mu,T_c)
\ C_{fp}(b|j_I;\mu^\prime,J^-_c,T_c^\prime) \nonumber\\
&&\phantom{\times}\times \big(\mu,J^-_\mu,T_c |\!|\!|\sum 
\frac{r_j^2}{R^2} \vec{\tau} (j) |\!|\!|\mu^\prime ,J^-_\mu, T_c^\prime \big)
\ \Bigg] \nonumber
\end{eqnarray}
as $J_{\mu^\prime}=J_\mu$.
As the $\widetilde{\mathcal{M}}_{2-b}^{(0)}$ operator transforms as a scalar
in isospin space, its matrix elements have the same sign in 
 both nuclei of the isospin doublet.

The parent state can have $T=0$ or 1, and in principle we have to consider
both diagonal and non-diagonal matrix elements (in the parent-state variables)
of the isovector 
operator $\sum (r_j^2/R^2)\ \vec{\tau}(j)$.
Obviously, its matrix elements vanish when $T$ or $T^\prime$ is equal to zero. 
Otherwise, we can use again a fractional parentage expansion. Only 
terms having the same parent can contribute to the matrix element and, in 
addition, the one-body operator has non-diagonal terms only between 
single-particle states
(with equal $j^\pi$) differing by at least {\em  two units} of the principal
quantum number: {\it i.e.}, it does not possess  {\em non-diagonal} matrix 
elements inside our
model space. 
As for the diagonal ones, 
shells (or sub-shells) completely filled with
protons and neutrons do not contribute to the sum, as 
they necessarily have $T=0$. 
If the valence nucleons are all in the same subshell (or, approximately, 
in subshells
with similar  $<(r_j/R)^2 >\approx <(r/R)^2 >_v$), the integral over the radial 
coordinates can be
factorised, $\sum \vec{\tau}(i)=2\vec{T}$,  and only the diagonal terms
with $\mu^\prime=\mu$ survive. Therefore, the matrix element takes the form
\begin{eqnarray}\label{E:62}
&&\hspace{-7mm}\big(\mu,J_\mu,T_c |\!|\!|\sum_j (r_j^2/R^2) \vec{\tau} (j) 
|\!|\!|\mu^\prime ,J_\mu, T_c^\prime \big)\\
&\approx& <(r/R)^2 >_v 
\ \big(J_\mu|\!|1|\!|J_\mu \big)\ \big(T_c |\!| 2\vec{T} 
|\!| T_c^\prime \big)\ \delta_{\mu,\mu^\prime} \nonumber
\end{eqnarray}
where $ <(r/R)^2 >_v $ is the average over active valence nucleons, and
 $\big(J_\mu|\!|1|\!|J_\mu \big)=\widehat{J}_\mu$.
For $T_c=T_c^\prime=1$,
$(T_c |\!| \vec{T} |\!| T_c )=\sqrt{T_c(T_c+1)(2T_c+1)}= \sqrt{6}$.
By comparing the result with Eq. (\ref{E:8}), we obtain approximately 
(as the first 6-J coefficient has the value 
$-1/3$):
\begin{eqnarray}\label{E:65}
&&\hspace*{-15mm}\Big( b_0;J_b;1/2,T_3 \Big|\!\Big| 
\widetilde{\mathcal{M}}_{2-b}^{(0)}
\Big|\!\Big| a_0;J_a,1/2,T_3 \Big)\nonumber\\*
&\approx&- C\ \frac{2}{3}\ \mathcal{A}(1) \frac{e}{2} \big( j_N 
|\!| r Y^{(1)}(\hat{r})  |\!| j_I \big)
\left< (r/R)^2\right >_v
\end{eqnarray}
Actually, the expectation values of $r^2/R^2$ for the different
orbitals of the $pf$ shell (estimated with Woods-Saxon wavefunctions)
do not differ more than $3\%$ from their average
value 0.615, as we obtain in Appendix \ref{S:a.2}.
By using this average value, one obtains
 for the numerical coefficient of the 2-body term
$(2/3)\left< (r/R)^2\right>_v =0.410$.
As this value is not negligible in comparison to that of the 1-body term 
(0.752), a sizable quenching of the isoscalar transition amplitude 
corresponding to the 1-body term results from the negative interference of 
the 2-body term.
A similar effect is found for the E1 transitions with $\Delta T=0$
in the $N=Z$ nuclei~\cite{bini}. 
However, in the present case the quenching only concerns the parent $T=1$ term.
As the parent $T=0$ term of Eq.~(\ref{E:54}) has no counterpart
in the 2-body matrix element, its contribution remains unaltered.

If we assume that the most important contribution to the asymmetry is due to 
the effect of coherent mixing, as approximated in this paragraph, 
 we obtain
\begin{eqnarray}
\label{E:66}
\epsilon (T_3)&\equiv& \frac{(b,7/2^-;1/2,T_3|\!|
\widetilde{\mathcal{M}}_E^{(10)}|\!|a,
9/2^+;1/2,T_3)}%
{(b,7/2^-;1/2,T_3|\!|\mathcal{M}_E^{(11)}|\!|a,9/2^+;1/2,T_3)}\\
&=&(-1)^{1/2+T_3}\ 3 C\ \frac {\left< j_N|(r/R)^3|j_I\right>}%
{\left< j_N|r/R|j_I\right>}
 \times \frac{\eta \mathcal{A}(1)-\mathcal{A}(0)}%
{ \mathcal{A}(1) + 3\mathcal{A}(0)} \ ,
\nonumber
\end{eqnarray}
where the quenching factor $\eta$ takes into account the negative interference 
with the two-body term of Eq.~(\ref{E:53}).

Equation (\ref{E:66}) only gives an approximate estimate of the effect, due to 
the many simplifying assumptions (notably, the closure approximation) 
that have been introduced to obtain this result. Moreover, inclusion in the 
model space of other orbitals of the upper major shell (as discussed in
Appendix \ref{S:a3}) would somewhat alter this result.
However, it could be instructive to evaluate some numerical results, also in
the limited space considered, to show that the coherent mixing with the 
IVGMR {\em can} explain the large values of the E1 asymmetries observe in our
example of the $A=67$ doublet, while the simplest processes discussed in the
previous sections were not able to do. 

With the above estimate,
$\eta=(0.752-0.410)/0.752=0.458$, and 
the asymmetry ratio for the mirror E1 strengths is
\begin{eqnarray}
\mathcal{R}&\equiv&\frac{B(E1,T_3=-1/2)}{B(E1,T_3=+1/2)}=
\left[\frac{1+\epsilon^-}{1-\epsilon^-} \right] ^2 \ ,
\end{eqnarray}
where we have put $\epsilon^-\equiv \epsilon(-1/2) =-\epsilon(+1/2)$.
Now, to obtain a more accurate estimate one should know the ratio 
$\mathcal{A}(0)/\mathcal{A}(1)$, which in turn depends on the $C_{fp}$
coefficients. 

The relative sign of $\mathcal{A}(0)$ and $\mathcal{A}(1)$
 depends on the 
combined effect of all terms in the sum of Eq.~(\ref{E:8}).
However, we can notice that each of them contains a factor
$(-1)^T$. If any of these terms 
dominates, the relative sign of 
$\mathcal{A}(0)$ and $\mathcal{A}(1)$ is well defined and
{\em negative}. 
Actually, this is very probably the case also under somewhat
broader conditions. Most probably, the second line of 
Eq.~(\ref{E:8}) (corresponding to negative--parity parents)
is only a small correction in comparison to the first one.
{
Let us consider, from now on, the numerical values 
corresponding to the $A=67$ doublet. We can note that
 the expression}
\begin{eqnarray}
(-1)^J\ \left\{ 
\begin{array}{c c c}
9/2&9/2&J\cr
7/2&7/2&1
\end{array}
\right\}\nonumber
\end{eqnarray}
has always the same (negative) sign for all $J$ values 
(from 0 to 7) and its value changes very slowly as long as
$J\leq 3$. Therefore, unless the parentage coefficients
have a very singular behaviour, the relative sign is determined
only by the factor $(-1)^T_c$ (see also Eq. (\ref{E:4})).

{
To obtain just an order-of-magnitude estimate of the expected 
effect, we could evaluate the asymmetry in the $A=67$ doublet, for two limiting
cases in which one of the two coefficients
  $\mathcal{A}(1)$ and $\mathcal{A}(0)$ is negligible in comparison 
to the other. 
Neglecting $\mathcal{A}(1)$ one obtains
$\epsilon^- \approx -0.753 C\approx -0.0872$ and the 
asymmetry ratio $\mathcal{R}\approx 0.705$.
 
Taking into account also $\mathcal{A}(1)$ would bring to smaller asymmetry
(larger $\mathcal{R}$) if $\mathcal{A}(1)$ and $\mathcal{A}(0)$
have the same sign, but can also result in a larger asymmetry
if -- as it is most probable -- they have opposite sign.
If, instead, $\mathcal{A}(0)$ is negligible 
in comparison to $\mathcal{A}(1)$, $\epsilon^-$ is
positive and its value depends on the coefficient $\eta$, which takes into
account the negative interference of the core terms. 
With $\eta=0.458$, for
$\mathcal{A}_0\ll\mathcal{A}_1$ one obtains
 $\epsilon^-\approx +0.120$ and $\mathcal{R}\approx 1.62$. 
Again, a larger asymmetry could be obtained  if also a contribution
from $\mathcal{A}_0$ (having opposite sign) is included.}

These results do not change appreciably if one assumes a charge distribution 
of Woods-Saxon shape (Appendix \ref{S:a.2}): one obtains
  $\eta=0.445$;
for $\mathcal{A}_1\ll \mathcal{A}_0$
 $\epsilon^- \approx -0.0852$
 and $\mathcal{R}\approx 0.710$; for  
$\mathcal{A}_1\gg \mathcal{A}_0$,  $\epsilon^- \approx 
+0.116$ and
$\mathcal{R}\approx=1.58$.

A last comment concerns the {\em expected sign} of $\epsilon^-$. 
If the dominant term
in the Eq.~(\ref{E:66}) is the one with $T=1$ parent, $\epsilon^->0$ and 
the reduced
strength should be larger in the nucleus with $N=Z+1$, for all transitions
between $g_{9/2}$ and $f_{7/2}$. The opposite is true if the $T=0$ parent
dominates.
 Again, qualitative considerations can help in predicting the relative
importance of the two terms. It is likely, in fact, that one of the most
important parents be the lowest $J=0$. Now, if $A=4n+1$, the lowest
$J=0$ parent state is the ground state of the even even 
self-conjugate nucleus with $A-1$ nucleons. Instead, if $A=4n-1$ (as in the
case $^{67}$As -- $^{67}$Se), the selfconjugate parent nucleus is odd-odd
and the lowest $J=0$ parent has $T=1$.
{
If this consideration is correct, the predicted sign of the
asymmetry is consistent with the experimental results in the $A=67$
mirror pair.}

\section{\label{S:d}Conclusions}

It seems worth summarising the results obtained for the different processes which 
could, in principle, produce an asymmetry in the E1 transition strength,
as observed in the case of the $^{67}$As $-\ ^{67}$Se mirror pair.
Higher-order terms, either of ``electric'' or ``magnetic'' origin, 
 usually excluded from calculations
by the approximation linked to the Siegert's theorem,
in the case considered are three orders of magnitude lower than the leading one. 
We note that these
corrections apply to the transition operator and not to the level
wavefunctions. Therefore, as long as -- as it was 
assumed here -- most of the shell-model terms contributing to the E1 
transition involve the same pair
of single-particle states, the same combination of fractional
parentage coefficients is involved for both the isoscalar and the isovector
term. {Thus if the isovector term} is hindered as a consequence of accidental
cancellation, a similar hindrance factor can be expected also to the isoscalar,
leaving the ratio almost unchanged. Only meson currents, neglected in our
approximate estimation of the magnetic term, could break, to some extent, the 
above conclusion.  

The Coulomb interaction, mixing in a different way the level wavefunctions 
in the two mirror nuclei, is presumably at the origin of the observed 
asymmetries. Its effect could be enhanced when a pair of levels having
equal $J^\pi$ lie, accidentally, close together.  
{
E.g.,} this could have been
the case for the two $7/2^-$ levels lying between 640 and 1100 keV in
$^{67}$As and $^{67}$Se. However, if { the asymmetry originated uniquely from the 
mixing between the two daughter} levels, the total sum of
the reduced strengths of the  E1 transitions feeding these levels
ought to be equal in the mirror nuclei, in contrast with the experimental evidence. 

The Coulomb mixing could also be enhanced if it took place between states with two 
``very similar'' wavefunctions. In Section~\ref{S:4.2} we  considered
an hypothetical mixing between a ``isospin analogue'' state and its
 corresponding ``antianalogue''. 
{
In the case of mass $A=67$, this} mixing {would lead to an asymmetry similar in 
size to the} observed effect. {It would also give} the right sign for the asymmetries.
However, this would only
happen if our $T=1/2,J^\pi=9/2^+$ state would be the exact antianalogue of the
lowest $T=3/2$ state with the same $J^\pi$, while some spread of the 
antianalogue strength among different levels is expected also in this region
of nuclei~\cite{Fodor76,Maripuu70}.

The effects of Coulomb mixing considered {thus far
only involved} states in the same Hilbert subspace needed to describe the parent
and daughter states of the E1 transition: in the simplest case, a full major
shell 
and at least one particle-hole 
excitation to the next
major shell.
A shell-model calculation in this Hilbert space could treat,
on the same footing, both the regular (isovector) part of the E1 transition 
amplitude and the ``induced-isoscalar'' term  originating from the mixing.
In such a calculation, the isovector part of the two-body Coulomb interaction
could be added to the empirical residual interactions, which could also
include the {symmetry-violating} part necessary to account for the Coulomb
Energy Differences~\cite{Lenzi}.

Finally, we have considered the possible effect of mixing with states
{\em outside} the truncated shell-model space, as those belonging to the
Giant Isovector Monopole Resonances. With the approximations discussed in
Section \ref{S:4.3}, { this effect could also be expressed in} a form that could
be treated in the truncated space, if the mean excitation energy of the
monopole resonance were at least approximately known. 

A shell-model calculation in such a restricted basis could therefore be able to
identify the origin of the observed asymmetry in E1 transition strengths.
At the moment, the coherent contribution of states belonging to the 
Giant Isovector Monopole resonance appears as the most probable candidate.

\section*{Acknowledgements}
One of us (P.G.B.) gratefully acknowledges Prof. B. Mosconi for useful 
discussions. {One of us (R.O.) gratefully acknowledges financial support from 
the Spanish Ministry of Economy and Competitiveness, via the Project Consolider 
Ingenio - CPAN - (CSD2007-42)}

\appendix
\section{\label{S:am}Evaluation of 
the reduced matrix elements for the magnetic term}

Here we  evaluate the reduced matrix elements of the operators
entering in the second line of Eq. (\ref{E:s1}), between single particle states
$\ell_1,j_1$ and $\ell_2,j_2$. To this purpose, 
 the following property \cite{talmi} of vector spherical harmonics  is exploited:
\begin{eqnarray}\label{E:m1}
\vec{\mathcal{Y}}^{(L,1)J}_M \cdot \vec{v}=
\left[ Y^{(L)} \otimes v^{(1)} \right]^{(J)}_M,
\end{eqnarray}
where $\vec{v}$ is a generic vector. {Here the cases
$\vec{v}=\vec{s}$ and $\vec{v}=\vec{j}$ are considered}.\\
In the first case, the reduced matrix element of the tensor
product can be obtained easily, because $s$ and $Y^{(L)}$ operate
on different Hilbert spaces 
\begin{eqnarray}\label{E:m2}
&&\hspace*{-7mm}\left( \ell_1,J_1 |\!| \left[ Y^{(L)} \otimes s^{(1)} 
\right]^{(J)}
 |\!| \ell_2, J_2\right)\\
&=&
\widehat{J}_1 \widehat{J} \widehat{J}_2
\left\{ \begin{array}{ccc}
\ell_1&1/2&J_1\cr
\ell_2&1/2&J_2\cr
L&1&J \end{array} \right\}
\left( \ell_1|\!|Y^{(L)} |\!| \ell_2\right)
\left( 1/2|\!|s^{(1)} |\!| 1/2\right)\nonumber
\end{eqnarray}
where $J=L$ and $\left( 1/2|\!|s^{(1)} |\!| 1/2\right) =\sqrt{3/2}$.
The relation
\begin{eqnarray}
&&\hspace*{-7mm}\left( \ell_1,J_1 |\!|  Y^{(L)}|\!| \ell_2, J_2\right)
=(-1)^{J_2+\ell_1+L+1/2}\\
&\times& \frac{\sqrt{3}}{2}
 \widehat{J}_1 \widehat{J}_2
\left\{ \begin{array}{c c c}
J_1&J_2&L\cr
\ell_2&\ell_1&1/2
\end{array}
\right\}
\left( \ell_1 |\!|  Y^{(L)}|\!| \ell_2 \right)
\nonumber
\end{eqnarray}
{can be exploited} to express the result in function of 
$\left( \ell_1,J_1 |\!|  Y^{(L)}|\!| \ell_2, J_2\right)$
as in Eq. (\ref{E:12}):
\begin{eqnarray}\label{E:m3}
&&\hspace*{-7mm}\left( \ell_1,J_1 |\!| \left[ Y^{(L)} \otimes s^{(1)} 
\right]^{(L)}
 |\!| \ell_2, J_2\right)\\
&=&
(-1)^{J_2+\ell_1+L+1/2}\ \sqrt{2}\ \widehat{L}
\ \left\{ \begin{array}{ccc}
\ell_1&1/2&J_1\cr
\ell_2&1/2&J_2\cr
L&1&L \end{array} \right\}\nonumber \\
&&\phantom{mm}\times  \left\{ \begin{array}{c c c}
J_1&J_2&L\cr
\ell_2&\ell_1&1/2
\end{array}
\right\}^{-1}
\left( \ell_1,J1|\!|Y^{(L)} |\!| \ell_2,J_2\right)
\nonumber
\end{eqnarray}

The second case is not so simple, because the operators $j$ and 
$Y^{(L)}$ do not commute, so that the symmetrised form
of the operator must be employed. Furthermore, they
 operate on the same Hilbert space, but one
can exploit the fact that $\vec{j}$ has no matrix elements
between different single-particle states to obtain:
\begin{eqnarray}\label{E:m4}
&&\hspace{-3mm}\frac{1}{2}\ \left( \ell_1,J_1 |\!| \left[ Y^{(L)} 
\otimes j^{(1)} \right]^{(J)}
+ \left[ j^{(1)}  \otimes Y^{(L)} \right]^{(J)}
 |\!| \ell_2, J_2\right)\nonumber \\
&=&\frac{1}{2} \left( \ell_1,J_1 |\!|  Y^{(L)}|\!| \ell_2, J_2\right)
\ (-1)^{J_1+J+J_2}\widehat{J} \nonumber\\
&\times&
\left[
\left\{ \begin{array}{ccc}
1&L&J\cr
J_2&J_1&J_2
\end{array} \right\}
\left( J_2 |\!| j^{(1)}|\!| J_2\right)\right.\\
&+&\left.
\left\{ \begin{array}{ccc}
L&1&J\cr
J_1&J_2&J_1
\end{array} \right\}
\left( J_1 |\!| j^{(1)}|\!| J_1\right)
\right] \nonumber
\end{eqnarray}
where $(j|\!|j^{(1)}|\!|j)=\sqrt{j(j+1)(2j+1)}$.

In the present case, $L=J=1,\ \ell_1=4,\ J_1=9/2$, $\ell_2=3,\ J_2=7/2$. 
With these numerical values, the coefficients of the reduced matrix element
of $ Y^{(L)}$
in the Eqs. (\ref{E:m3}, \ref{E:m4}). are, respectively, $\sqrt{1/6}$ and
$-\sqrt{1/2}$.

\section{\label{S:a.2}Radial wavefunctions and Coulomb potential with 
a Woods-Saxon distribution}
The radial wavefunctions have been calculated assuming a Woods-Saxon
potential plus spin-orbit:
\begin{eqnarray}
V(r)&=& \left[V_0+V_s \vec{\ell}\cdot \vec{s}
\ \frac{r_0^2}{r}\ \frac{\rm d}{\rm dr} \right] \ \frac{1}{1+ e^{(r-R_0)/a}}
\end{eqnarray}
with the values of the constants consistent with Bohr and Mottelson
\cite{BM}: $V_0=-51$ MeV, $V_s=22$ MeV, $R_0=r_0~A^{(1/3)}$, $r_0=1.27$ fm  
and $a=0.67$ fm.
\begin{figure}
\includegraphics[height=12cm,bb=40 15 475 720]
{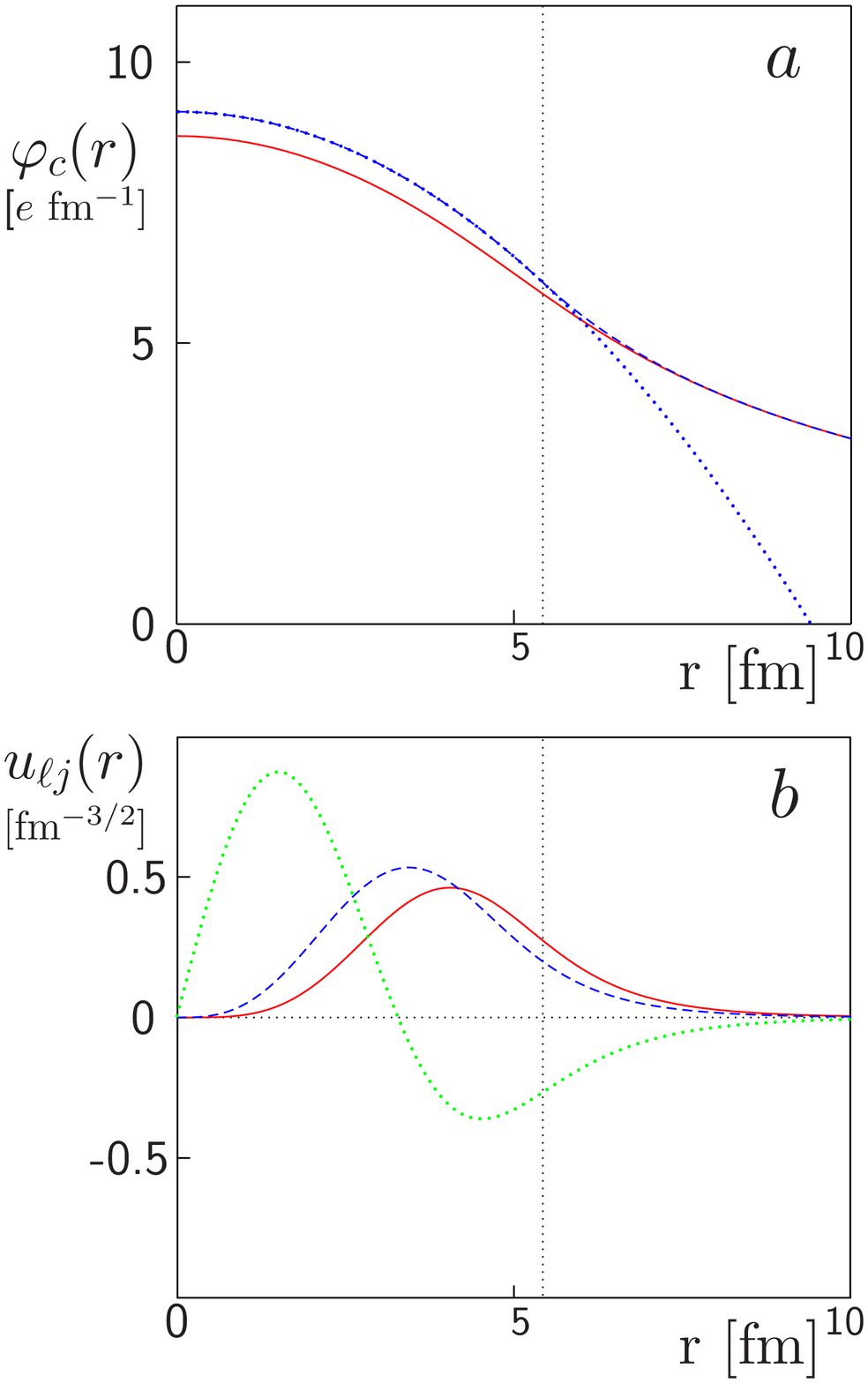}
\caption{\label{F:a2}(color online) $a$: electrostatic potential for 
uniformly charged
sphere (dashed) and with a Woods-Saxon charge distribution for $A=67$
 (continuous line).
The dotted line shows the continuation outside the sphere of the 
expression for the uniform distribution (dashed curve) in the internal region.
$b$:~examples of radial wavefunctions for the Woods-Saxon potential 
($+$ spin-orbit) with
the parameters suggested in \cite{BM}: $0g_{9/2}$ (continuous line),
$0f_{5/2}$ (dashed), $1p_{1/2}$ (dotted).
The vertical dotted line corresponds to the value of the nuclear radius $R$.
}
\end{figure}

For a consistent evaluation of Coulomb interactions,
one needs the average electrostatic potential  $\varphi_c(r)$ 
of a distribution of $A-1$ point
charges $e/2$, which will be approximated with a continuous charge 
distribution having a Woods-Saxon shape:
\begin{eqnarray}
\rho_e(r)= \frac{\rho_0}{1+e^\frac{r-R_0}{a}}
\end{eqnarray}
where 
\begin{eqnarray}
\rho_0= (A-1)\frac{e}{2}
\ \left[ \int \frac{1}{1+e^\frac{r-R_0}{a}}\ 4\pi r^2 {\rm d}r\right]^{-1}
\end{eqnarray}

With the condition that $ \varphi_c(r)\to 0$ for
$r\to \infty$, we obtain
\begin{eqnarray}
 \varphi_c(r)\equiv  \rho_0  
=\int_r^\infty \frac{{\rm d}y}{y^2}\int_0^y 
\frac{\rho_0}{1+e^\frac{x-R_0}{a}}\ x^2
\ {\rm d}x
\end{eqnarray}
This integral has been evaluated numerically, for $A=67$, with the 
parameter values suggested in \cite{BM}. In fig.\ref{F:a2},
the result is compared with the potential of an uniformly charged sphere
of charge density equal to $\rho_0$ and total charge $(A-1)e/2$. 
The radius $R$ of
the sphere is determined by the condition
\begin{eqnarray}
\frac{4\pi}{3}R^3=\frac{(A-1)e}{2\rho_0}=\int_0^\infty 
\frac{1}{1+e^\frac{r-R}{a}}\ 4\pi r^2 {\rm d}r
\end{eqnarray}
 To simplify the comparison of the  results,
$\varphi_c(r)$ {is expressed} in terms
of the adimensional function $f_c(r/R)$:
\begin{eqnarray}
\varphi_c(r) \equiv  \frac{(A-1)e}{4R} f_c(r/R)\ ,
\end{eqnarray}
and we define $\Delta f_c(r)=f_c(0)-f_c(r)$. For the (extrapolated) potential
of the uniformly-charged sphere, one obtains $\Delta f_c(r)=(r/R)^2$.
One must now calculate the matrix elements of the operators 
$\widetilde{\mathcal{M}}_{1-b}^{(0)}$ and $\widetilde{\mathcal{M}}_{2-b}^{(0)}$
defined in Section \ref{S:4.3}. For the one-body term, we consider the ratio
\begin{eqnarray}
\frac{ \left< f_{7/2} | (r/R) f_c(r/R) |g_{9/2} \right>}
{ \left< f_{7/2} | r/R |g_{9/2} \right>}
\end{eqnarray}
while for the two-body term (and also for the calculations of 
Section \ref{S:4.2}), it is sufficient to evaluate the diagonal
matrix elements of $f_c(r/R)$

By  numerical integration, with the parameters of \cite{BM} one
obtains the values of the necessary integrals reported in the last 
column of Table \ref{tab:aa1}. In the other columns, the corresponding
values are calculated, with the Woods-Saxon wavefunctions, 
for the potential of the uniformly
charged sphere and for the extrapolation of the inner potential outside
the sphere (dotted line in Fig. \ref{F:a2}$a$). 
\begin{table}
\caption{Values of radial integrals 
for different assumptions on the Coulomb potential. In all cases,
$< f_{7/2}|r/R |g_{9/2} > = 0.8285 $.}
\label{tab:aa1}
\begin{tabular}{cccc}
\hline\noalign{\smallskip}
& \multicolumn{2}{c}{Constant $\rho_c$} & 
Woods-Saxon  \\
&  sphere & extrapol. & 
distribution \\
\noalign{\smallskip}\hline\noalign{\smallskip}
{\large $\frac{<f_{7/2}|r\Delta f_c|g_{9/2}>}{<f_{7/2}|r|g_{9/2}>}$}
&0.700 &0.752 & 0.739\\
$<g_{9/2}| \Delta f_c| g_{9/2}> $& 0.697 & 0.749 & 0.735 \\
$<f_{7/2}| \Delta f_c| f_{7/2}> $& 0.594 & 0.625 & 0.620 \\
$<f_{5/2}| \Delta f_c| f_{5/2}> $& 0.564 & 0.592 & 0.587 \\
$<p_{3/2}| \Delta f_c| p_{3/2}> $& 0.572 & 0.625 & 0.608 \\
$<p_{1/2}| \Delta f_c| p_{1/2}> $& 0.580 & 0.636 & 0.617 \\
\noalign{\smallskip}\hline
\end{tabular}
\end{table}

\section{\label{S:a3}Effect of the inclusion of more orbitals}
Until now, we have assumed that only the intruder orbit $g_{9/2}$ is significant
for the description of the relevant states. As a consequence, only the
transitions between $g _{9/2}$ and $f_{7/2}$ contribute to E1.
If other orbitals of the upper major shell ({\it e.g. $1d_{5/2}$}) are taken
into account, other orbitals of the lower major shell can be involved
in the E1 transitions.
We consider now the changes that must be introduced 
in our calculations as a consequence 
of the inclusion in the model space of the two complete major shells. 

Equation (\ref{E:1}) must be modified as follows:
\begin{eqnarray}
\label{E:a3.1}
|a;J_a,M_a;1/2,T_3> &=&\\
&&\hspace*{-38mm}\phantom{+}\sum_{j_I}\sum_\mu C_{fp}(a|j_I;\mu,J_\mu^+,T_c)
 [\phi(j_I) \otimes \Phi(\mu,J_\mu^+,T_c)]^{(J_a,1/2)}_{M_a,T_3} \nonumber \\
&&\hspace*{-43mm}\phantom{\times \Big[ }+\sum_{j_N} 
\sum_\mu C_{fp}(a|j_N;\mu,J_\mu^-,T_c)  
[\phi(j_N) \otimes \Phi(\mu,J_\mu^-,T_c)]^{(J_a,1/2)}_{M_a,T_3}\nonumber
\end{eqnarray}
and similarly  Eq.~(\ref{E:2}).
Equation~(\ref{E:6}) becomes
\begin{eqnarray}\label{E:a3.6}
&&(b,J_b;T_b,T_3|\!|\mathcal{M}_E^{(1K)}|\!|a,J_a;T_a,T_3)= (-1)^{1/2-T_3}
\nonumber \\
&\times&
\left( \begin{array}{c c c}
T_b&K&T_a\cr
-T_3&0&T_3
\end{array}\right) 
\sum_{j_I,j_N} (j_N|\!|\!|\mathcal{M}_E^{(1K)}|\!|\!|j_I)
\nonumber\\
&&\times  (-1)^K \sum_{T_c=0,1} \mathcal{A}_{j_I,j_N} (T_c)
\left\{ \begin{array}{c c c} 
1/2&T_b&T_c\cr
T_a&1/2&K
\end{array}\right\} 
\end{eqnarray}
with $ \mathcal{A}_{j_I,j_N} (T_c)$ given by Eq.~(\ref{E:8}).
Finally, Eqs. (\ref{E:10},\ref{E:11}) become
\begin{eqnarray}
\label{E:a3.10}
&&(b,J_b;1/2,T_3|\!|\mathcal{M}_E^{(11)}|\!|a,J_a;1/2,T_3)
=\frac{(-1)^{1/2+T_3}}{6}\\
&&\phantom{mmm}\times 
 \sum_{j_I,j_N}
\left[ \mathcal{A}_{j_I,j_N}(1) + 3\mathcal{A}_{j_I,j_N}(0)
\right]  (j_I|\!|D^{(1)}_{IV}|\!|j_N)\nonumber\\
\label{E:a3.11}
&&(b,J_b;1/2,T_3|\!|\mathcal{M}_E^{(10)}|\!|a,J_a;1/2,T_3)\\
&&\phantom{mmm}=\frac{1}{2} \sum_{j_I,j_N}
\left[ \mathcal{A}_{j_I,j_N}(1)-\mathcal{A}_{j_I,j_N}(0) \right]   
(j_I|\!|D^{(1)}_{IS}|\!|j_N)
\ .\nonumber
\end{eqnarray}
With these modifications the possible consequences of the inclusion of more 
orbitals on the results of the different sections can now be considered.

Section \ref{S:2} only concerns the form of the E1 {\em operator}, and does
not depend on the assumed form of the wavefunctions.

Section \ref{S:4.1} also is completely valid, as the considerations reported
there do not depend on the details of the wavefunctions.

Section \ref{S:4.2} depends on the assumed structure of the analogue and
anti-analogue states. The choice given there presumably corresponds to
an upper limit of the mixing. For example, in Eq. (\ref{E:44}), the choice of 
a pure $g_{9/2}$ orbit corresponds to the maximum possible value of
the expectation value of $r^2/R^2$. Our conclusion, {\it i.e.} that this process is
not able to explain the observed effect, is therefore even stronger if
other orbitals are considered.

It remains to consider Section \ref{S:4.3}. 
The sum on $j_I,j_N$ must be included in Eqs. (\ref{E:54}, \ref{E:65})
to obtain the one-body and the two-body contributions to the induced isoscalar
E1:
\begin{eqnarray}
\label{E:a3.54}
&&\hspace*{-1mm}\Big(b,J_b;\frac{1}{2},T_3\Big|\!\Big|
\widetilde{\mathcal{M}}_{1-b}^{(0)}
\Big|\!\Big|a,J_a;\frac{1}{2},T_3 \Big)\nonumber\\
&& \hspace*{-1mm}= \frac{1}{2} \sum_{j_I,j_N}
\left[ \mathcal{A}_{j_I,j_N}(1)
-\mathcal{A}_{j_I,j_N}(0) \right]  
\frac{e}{2}\ \Big( j_N\Big|\!\Big| r\  Y^{(1)}(\hat{r})
\Big|\!\Big|j_I\Big)\nonumber \\
&& \hspace*{15mm}\times
\frac{\left< j_N|(r/R)^3|j_I \right> }{\left< j_N|(r/R)|j_I \right> }
\end{eqnarray}
and
\begin{eqnarray}\label{E:a3.65}
&&\hspace*{-1mm}\Big( b_0;J_b;1/2,T_3 \Big|\!\Big| 
\widetilde{\mathcal{M}}_{2-b}^{(0)}
\Big|\!\Big| a_0;J_a,1/2,T_3 \Big)\\*
&\approx&- C\ \frac{2}{3}\ \sum_{j_I,j_N}
\mathcal{A}_{j_I,j_N}(1) \frac{e}{2} \big( j_N 
|\!| r Y^{(1)}(\hat{r})  |\!| j_I \big)
\left< (r/R)^2\right >_v \nonumber
\end{eqnarray}
We obtain therefore
\begin{eqnarray}
\label{E:a3.66}
\epsilon (T_3)&\equiv& \frac{(b,7/2^-;1/2,T_3|\!|
\widetilde{\mathcal{M}}_E^{(10)}|\!|a,
9/2^+;1/2,T_3)}%
{(b,7/2^-;1/2,T_3|\!|\mathcal{M}_E^{(11)}|\!|a,9/2^+;1/2,T_3)}\\
&=&(-1)^{1/2+T_3}\ 3 C\ \times \cr
&&\hspace*{-10mm} \frac{
\sum_{j_I,j_N}
\left< j_N|(r/R)^3|j_I\right>%
\left[\eta_{j_I,j_N} \mathcal{A}_{j_I,j_N}(1)-\mathcal{A}_{j_I,j_N}(0)\right]}%
{\sum_{j_I,j_N}\left< j_N|r/R|j_I\right>
\left[ \mathcal{A}_{j_I,j_N}(1) + 3\mathcal{A_{j_I,j_N}}(0)\right]} \nonumber
\end{eqnarray}
where $\eta_{j_I,j_N}$ has the same meaning as in Eq.~(\ref{E:66}).

\end{document}